\documentclass[prl,twocolumn,showpacs,amsmath,showkeys,floatfix]{revtex4} 

\usepackage{amsmath}
\usepackage{graphicx}
\usepackage{color}

\def\gtorder{\mathrel{\raise.3ex\hbox{$>$}\mkern-14mu
	\lower0.6ex\hbox{$\sim$}}}
\def\ltorder{\mathrel{\raise.3ex\hbox{$<$}\mkern-14mu
	\lower0.6ex\hbox{$\sim$}}}

\def\lsim{\mathrel{\rlap{\lower4pt\hbox{\hskip1pt$\sim$}}
    \raise1pt\hbox{$<$}}}
\def\gsim{\mathrel{\rlap{\lower4pt\hbox{\hskip1pt$\sim$}}
    \raise1pt\hbox{$>$}}}

\def \beqn{\begin{eqnarray}}
\def \eeqn{\end{eqnarray}}

\def \bea{\begin{eqnarray}}
\def \beq{\begin{equation}}
\def \eea{\end{eqnarray}}
\def \eeq{\end{equation}}
\def \nn{\nonumber}
\def \bwt{\begin{widetext}}
\def \ewt{\end{widetext}}

\begin{document}
\title{New sum rule for the nuclear magnetic polarizability}

\pacs{11.55.Hx, 25.20.Dc, 25.30.Fj, 13.60.Fz}

\keywords{dispersion relations, Compton scattering, sum rules}

\author{Mikhail Gorchtein}
\affiliation{PRISMA Cluster of Excellence, Institut f\"ur Kernphysik, Johannes Gutenberg-Universit\"at, Mainz, Germany}
\email{gorshtey@kph.uni-mainz.de}

\begin{abstract}
I extend the well-known photonuclear sum rule that relates the strength of the photoexcitation of the giant dipole resonance in a nucleus to the number of elementary scatterers-nucleons to the case of virtual photons. The new sum rule relates the size of the magnetic polarizability of a nucleus to the slope of the transverse virtual photoabsorption cross section integrated over the energy in the nuclear range. I check this sum rule for the deuteron where necessary data is available, discuss possible applications and connection with other sum rules postulated in the literature.
\end{abstract}
\date{\today}

\maketitle


Scattering of light off a composite object has long been used to study its structure. At low frequencies, electromagnetic waves scatter without absorption and solely probe its mass and electric charge, the classical Thomson result. With the photon energy raising  above the absorption threshold internal structure is revealed. Kramers and Kronig related the photoabsorption spectrum of a material to its index of refraction by means of a dispersion relation \cite{Kronig,Kramers} based on the probability conservation and causality. Dispersion relations and sum rules have been among the main tools for studying the electromagnetic interactions in atomic, nuclear and hadronic physics domains. These domains roughly correspond to keV, MeV and GeV photon energies, respectively, and this scale hierarchy indicates that dynamics in each domain can be clearly identified. Thomas-Reiche-Kuhn sum rule equated the sum of oscillator strengths in an atom to the number of electrons \cite{Thomas,Reiche,Kuhn}. For nuclei, Levinger-Bethe \cite{Levinger} and Gell-Mann, Goldberger and Thirring \cite{GellMann:1954db} related the integrated photoabsorption cross section to the number of elementary scatterers, protons and neutrons in a nucleus. For GeV energy photons that resolve the nucleon structure, Gorchtein, Hobbs, Londergan and Szczepaniak \cite{Gorchtein:2011xx} observed that the integrated strength of the nucleon resonances may be explained by counting the constituent quarks. These sum rules are an economic, albeit approximate way to express duality, the transcendence of higher energy degrees of freedom in the low-energy phenomena \cite{Bloom:1970xb}. In this letter I extend the Thomas-Reiche-Kuhn-Levinger-Bethe sum rule to the case of virtual photons, obtain a sum rule for the nuclear magnetic polarizability, and discuss further applications. \\
\indent
The spin-averaged, forward Compton tensor $T^{\mu\nu}$ is expressed in terms of two scalar amplitudes $T_{1,2}(\nu,Q^2)$,
\beqn
T^{\mu\nu}&=&T_{1}(\nu,Q^2)\left(-g^{\mu\nu}+\frac{q^\mu q^\nu}{q^2}\right)\\
&+&T_{2}(\nu,Q^2)\frac{1}{M_T^2}
\left(p-\frac{(p\cdot q)}{q^2}q\right)^\mu\left(p-\frac{(p\cdot q)}{q^2}q\right)^\nu,\nn
\eeqn
with the invariants defined in terms of the nucleus and photon four-momenta $p,q$ as $\nu=(p\cdot q)/M_T$, $Q^2=-q^\mu q_\mu=-q^2\geq0$, and $p^2=M_T^2$, with $M_T$ the target nucleus mass. In this letter I concentrate on the transverse amplitude $T_1$. Its imaginary parts is related to the unpolarized structure function $F_{1}$ as 
Im$T_{1}=(\pi\alpha_{em}/M_T)F_{1}$, 
with $\alpha_{em}\approx1/137$ the fine structure constant. $T_1$ satisfies a once subtracted dispersion relation (DR), 
\begin{eqnarray}
{\rm Re}\,T_1(\nu,Q^2)=T_1(0,Q^2)
+\frac{\alpha_{em}\nu^2}{M_T} \int\limits_{0}^\infty\frac{d\nu'^2F_1(\nu',Q^2)
}{\nu'^2(\nu'^2-\nu^2)}\label{eq:DR1}
\end{eqnarray} 
where the integral is understood in terms of its  principal value. I remove the pole contribution that is due to an absorption of a virtual photon by an on-shell ground state (this separation is well-defined, see, {\it e.g.}, discussion in \cite{Birse:2012eb}). Upon this removal, the subtraction constant $T_1^{np}(0,Q^2)$ is defined in terms of the nuclear charge form factor $F_C$ normalized to unity at $Q^2=0$, and the nuclear magnetic polarizability $\beta_M^{nucl}(Q^2)$ generalized to finite $Q^2$,
\beqn
T_1^{np}(0,Q^2)=-\frac{\alpha_{em}}{M}\frac{Z^2F_C^2(Q^2)}{Z+N}+Q^2\beta^{nucl}_M(Q^2),\label{eq:LEX1}
\eeqn
with $Z(N)$ the number of protons (neutrons) in the nucleus, $\alpha_{em}\approx1/137$ the fine structure constant, $M\approx M_p\approx M_n$ the nucleon mass, such that $M_T\approx(Z+N)M$.\\
\begin{figure}[h]
\includegraphics[width=8.5cm]{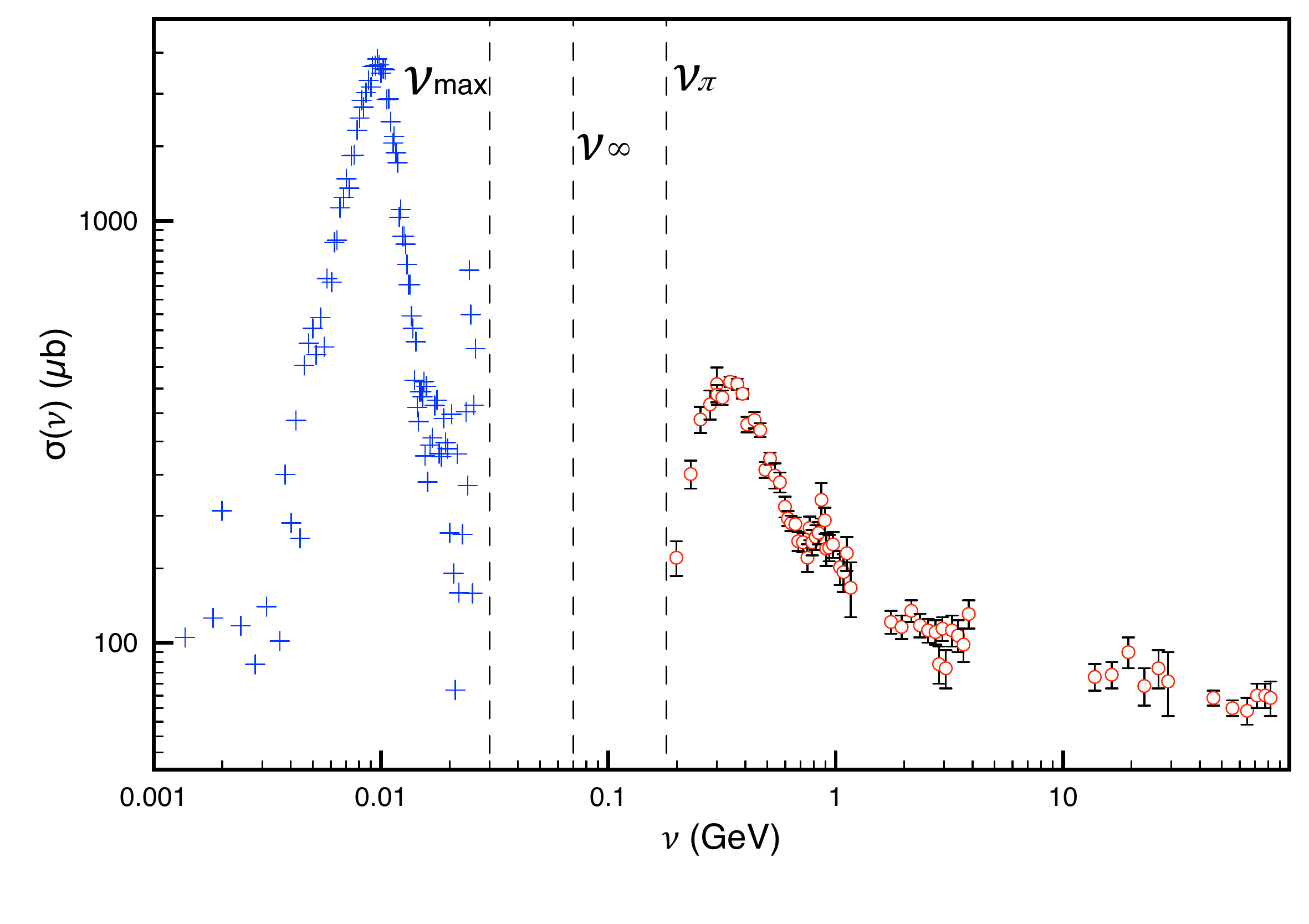}
\caption{(Color online) Total photoabsorption cross section on lead in $\mu$barn as function of energy. Data in the nuclear range (blue crosses) extend up to $\nu_{max}\approx30$ MeV are from Ref. \cite{Harvey:1964}. Data above the pion production threshold $\nu_\pi$ (red open circles) are from \cite{Hesse:1970cy,Caldwell:1973bu,Caldwell:1978ik,Bianchi:1995vb}. The vertical dashed lines display $\nu_{max},\nu_\infty,$ and $\nu_\pi$, see text for further details.}
\label{fig:NuclSigma}
\end{figure}
Real photoabsorption on lead, shown in Fig. \ref{fig:NuclSigma}, illustrates several general features common to all nuclei: i) the strength of nuclear excitations is concentrated in the region between the breakup threshold $\nu_{min}(Q^2)=B+Q^2/(2M_T)$, with $B$ the nucleon removal threshold for the nucleus, and $\nu_{max}(Q^2)\approx B+Q^2/(2M)+30$ MeV; ii) nuclear cross sections stay small above that energy and below the threshold for the nucleon breakup $\nu_\pi(Q^2)=Q^2/(2M)+m_\pi+m_\pi^2/(2M)$, with $m_\pi$ the pion mass; iii) above this threshold, an incoherent absorption by $Z$ protons and $N$ neutrons that make up a nucleus is a good overall representation of the cross section (modulo nuclear effects). I exploit the observed gap between $\nu_{max}$ and $\nu_\pi$ by evaluating the DR for $T_1$ at an intermediate energy $\nu_{\infty}(Q^2)\approx B+Q^2/(2M)+70$ MeV, impose the hierarchy of scales, $\nu^2_{max}\ll\nu^2_\infty\ll\nu^2_\pi$ and take respective limits,
\beqn
&&\!\!\!\!{\rm Re}\,T^{np}_1(\nu_\infty,Q^2)=T^{np}_1(0,Q^2)-\frac{2\alpha_{em}}{M_T}\!\!\!\!\int\limits_{\nu_{min}}^{\nu_{max}}\!\!\frac{d\nu}{\nu}F_1(\nu,Q^2)\nn\\
&&\!\!\!\!+\frac{2\alpha_{em}\nu_\infty^2}{M_T} \int\limits_{\nu_\pi}^\infty\frac{d\nu}{\nu^3}F_1(\nu,Q^2)
+\frac{\alpha_{em}}{M_T}{\cal{P}}\!\!\!\!\!\int\limits_{\nu_{max}}^{\nu_{\pi}}\!\!
\frac{d\nu^2\nu_\infty^2F_1(\nu,Q^2)}{\nu^2(\nu^2-\nu_\infty^2)}.\nn\\
\label{eq:LEX_int}
\eeqn
\indent
For compactness, I suppressed the $Q^2$-dependence of the integration limits. The integral between $\nu_{max}$ and $\nu_\pi$ is understood in the sense of its principal value. Next, the scale hierarchy is used to calculate Re$\,T^{np}_1(\nu_\infty,Q^2)$: the scale $\nu_\infty$ was chosen such that the bulk of nuclear excitations lies significantly below it. Then, photons will scatter off essentially unbound nucleons; the energy is significantly lower than the pion production threshold, so the nucleon structure is not resolved at that energy, and it is legitimate to approximate its value by a low-energy expansion up to order $\nu_\infty^2$,
\beqn
&&\!\!\!\!\!\!{\rm Re}T_1^{np}(\nu_\infty,Q^2)=-Z\frac{\alpha_{em}}{M}F_{D}^{p\,2}(Q^2)-N\frac{\alpha_{em}}{M}F_{D}^{n\,2}(Q^2)\nn\\
&&+ZQ^2\beta_M^p(Q^2)+NQ^2\beta_M^n(Q^2)\nn\\
&&+\frac{2\alpha_{em}\nu_\infty^2}{M}\int_{\nu_\pi}^\infty\frac{d\nu}{\nu^3}\left[ZF_1^{p}(\nu,Q^2)+NF_1^{n}(\nu,Q^2)\right],
\label{eq:LEX2}
\eeqn
where $F_D^{p(n)}$ denotes the proton (neutron) Dirac form factor, and $\beta_M^{p(n)}(Q^2)$ stand for the proton (neutron) magnetic polarizability, respectively, extended to finite $Q^2$. A subtracted dispersion relation analogous to that of Eq. (\ref{eq:DR1}) is imposed on the single nucleon amplitudes, with $F_1^{p,n}$ free nucleon structure functions. Now, Eqs. (\ref{eq:LEX1},\ref{eq:LEX_int},\ref{eq:LEX2}) can be combined together, and the coefficients at different powers of $\nu_\infty$ equated. If nuclear and hadronic scales are indeed well-separated, above $\nu_{max}(Q^2)$ nucleons are unbound, and the coefficient at $\nu_\infty^2$ should vanish,
\beqn
&&\int_{\nu_\pi}^\infty\!\frac{d\nu}{\nu^3}\left[\frac{M}{M_T}F_1(\nu,Q^2)-ZF_1^{p}(\nu,Q^2)-NF_1^{n}(\nu,Q^2)\right]\nn\\
&&+\frac{M}{M_T}{\cal{P}}\int_{\nu_{max}}^{\nu_{\pi}}
\frac{d\nu F_1(\nu,Q^2)}{\nu(\nu^2-\nu_\infty^2)}=0. 
\label{eq:shadowing}
\eeqn
\indent
Turning to the terms independent of $\nu_\infty^2$, 
and setting $Q^2=0$ Levinger and Bethe \cite{Levinger} obtained, 
\beqn
ZN=2 \int_{\nu_{min}}^{\nu_{max}}\frac{d\nu
}{\nu} F_1(\nu,0),\label{eq:TRK}
\eeqn 
{\it i.e.} integrated strength of nuclear excitations is fixed by the number of nucleons within the nucleus. 
Levinger-Bethe sum rule of Eq. (\ref{eq:TRK}) is obeyed for a wide range of nuclei, typically better than 10\% \cite{BermanFultz1975}. As an example, the parametrization of the deuteron photodesintegration cross section in Ref. \cite{Carlson:2013xea} leads to the value of the right hand side 1.007, in excellent agreement with the sum rule, $NZ=1$. Deviations due to non-vanishing of the principal value integral and effects of nuclear binding and shadowing in Eq. (\ref{eq:shadowing}) were estimated, {\it e.g.}, in Refs. \cite{Levinger,GellMann:1954db}. 

I now consider the first derivative with respect to $Q^2$ at the origin. Using the charge radius defined as $R_{Ch}^2=-6F'_C(0)$, the sum rule for the nuclear magnetic polarizability is obtained, 
\beqn
\beta^{nucl}_M&=&\frac{2\alpha_{em}}{M}\int_{\nu_{thr}}^{\nu_{max}}\frac{d\nu}{\nu} \frac{d}{dQ^2}\left.F_1(\nu,Q^2)\right|_{Q^2=0}\nn\\
&-&\frac{Z^2\alpha_{em}}{(Z+N)M}\frac{R_{Ch}^2}{3},\label{eq:SR}
 \eeqn
where I neglected effects of nuclear and nucleon recoil that enter the $Q^2$-dependence of the integration limits (above taken at $Q^2=0$), effects of nucleon polarizabilities and nucleon charge radii. 

This sum rule is useful since for most nuclei the magnetic polarizability is not known, unlike the sum $\alpha_E^{nucl}+\beta_M^{nucl}$ that is fixed by Baldin sum rule \cite{Baldin},
\beqn
\alpha^{nucl}_E+\beta^{nucl}_M=\frac{2\alpha_{em}}{M_T}\int_{\nu_{min}}^\infty\frac{d\nu
}{\nu^3}F_1(\nu,0),
\eeqn
and can be directly extracted from the experimental data. 

To my knowledge, deuteron is the only nucleus for which theoretical predictions of $\beta_M^{nucl}$ exist, calculated in EFT \cite{Chen:1998vi} and potential model \cite{Friar} approaches, summarized as $\beta^d_M=0.072(5)$ fm$^3$.  One can now check, how important the neglected terms are numerically. Using the value of the proton charge radius from recent $\mu H$ measurements \cite{Pohl:2010zza,Antognini:1900ns}, and the neutron charge radius along with the nucleon magnetic polarizabilities from the PDG \cite{PDG} gives $\sim1.6\times10^{-3}$ fm$^3$, two orders of magnitude below $\beta_M^d$. The effect of the deuteron charge radius taken from \cite{Mohr:2012tt} is of the similar order, $\sim-1.5\times10^{-3}$ fm$^3$, also negligible. However, for heavy nuclei these two contributions can have very different size, {\it e.g.}, for lead the two terms give $\sim0.08$ fm$^3$ and $\sim-0.5$ fm$^3$, respectively, which explains the choice of keeping the nuclear radius effect but neglecting the nucleonic contributions. The value of $\beta_M$ for lead is unknown, but  $\alpha_E+\beta_M\approx14.5$ fm$^3$ \cite{{BermanFultz1975}} gives a rough idea, even though it can be expected that $\beta_M\lesssim0.1\alpha_E$ for that nucleus. 

\begin{figure}[h]
\includegraphics[width=8cm]{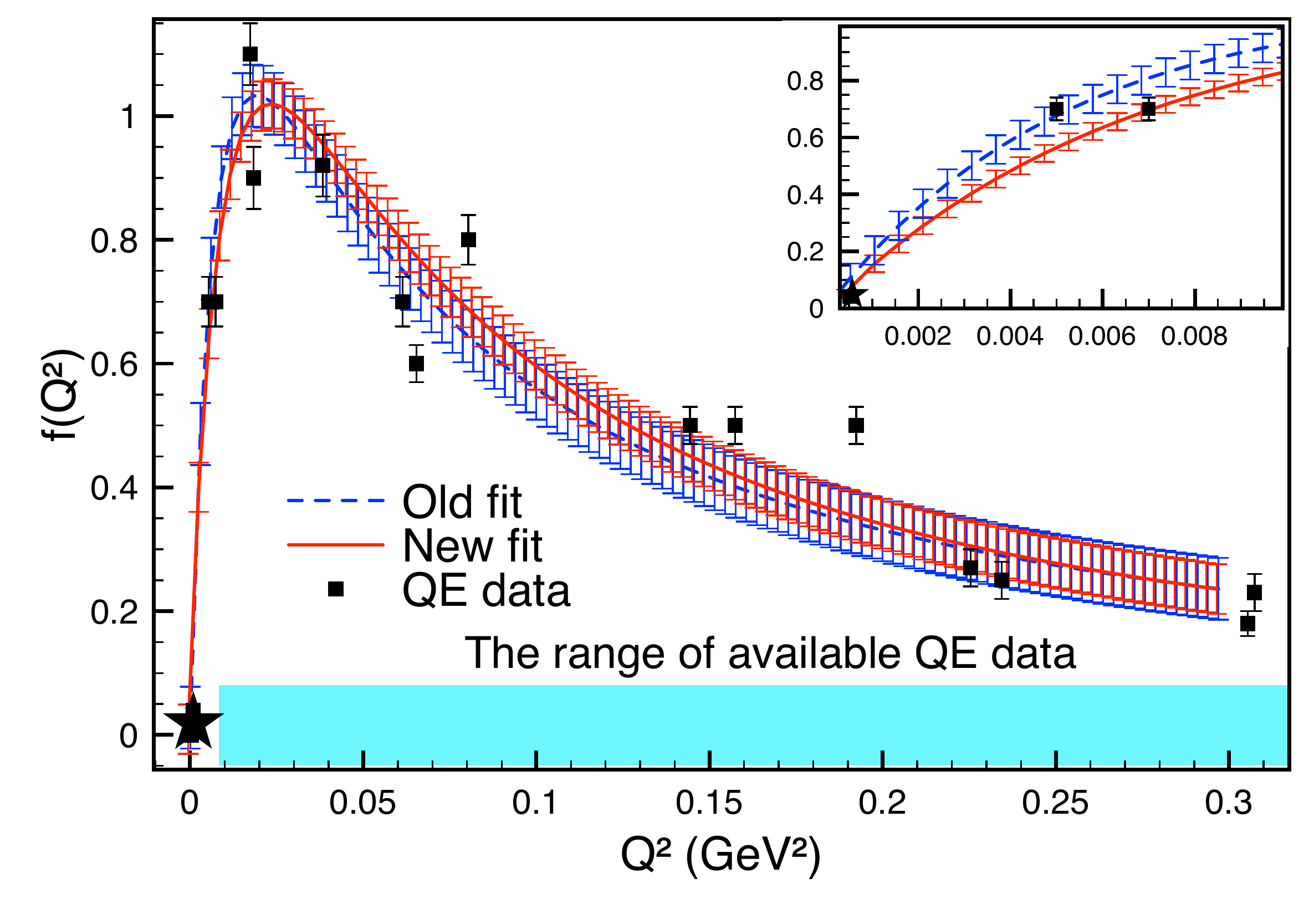}
\caption{(Color online) The comparison of the old fit without the use of the sum rule, $f_T^{FSI}$ (blue dashed curve) and the new fit using the sum rule, $\tilde f_T^{FSI}$ (red solid curve), with the uncertainty of each fit indicated by the band of the respective color. The sum rule is indicated by the star. The shaded band shows the kinematical range that is covered by the existing $D(e,e')pn$ data. The subview in the upper right corner magnifies the small values of $Q^2$ where the the slope of the new fit function is fixed to reproduce the value of $\beta_M^d$. Data points correspond to experimental data sets analyzed in  Ref. \cite{Carlson:2013xea} (Refs. [35-42] of that article).}
\label{fig:Fitfunctions}
\end{figure}

Using a recently proposed detailed parametrization of deuteron breakup data \cite{Carlson:2013xea} that covers $Q^2$ in the range [0.005 GeV$^2;\,3$ GeV$^2$] and energy between the deuteron breakup threshold and well into the hadronic range, a numerical evaluation of the right hand side of Eq. (\ref{eq:SR}) can be done. It leads to 
$\beta^d_M=0.096(15)\,{\rm fm}^3$,
close to the model-based expectation, $\beta^d_M=0.072(5)$ fm$^3$.  Note that even raising $\nu_{max}$ to 140 MeV would increase the integral by mere 1\%, so the result is very robust. To enforce the agreement, one needs to modify the parametrization of Ref. \cite{Carlson:2013xea} (Eq. (27) and Table II of  that Ref.) via 
\beqn
f_T^{FSI}(Q^2)=\frac{2.15(35)\times10^4\,{\rm GeV}^{-3}Q^2}{(1+52(8)\,{\rm GeV}^{-2}Q^2 )^2}
\eeqn
to
\beqn
\tilde f_T^{FSI}(Q^2)=\frac{1.61(11)\times10^4\,{\rm GeV}^{-3}Q^2}{(1+35(6)\,{\rm GeV}^{-2}Q^2 )^{2.2}}.\label{eq:ftilde}
\eeqn
\indent
The error in the numerator is fixed by that in the value of $\beta_M^d$, and the error (and a different power) in the numerator is obtained by a new fit to the quasi elastic data, as described in Ref. \cite{Carlson:2013xea}. The two fit functions are shown in Fig. \ref{fig:Fitfunctions}.
With this exercise I demonstrate that the existing deuteron quasi elastic data are consistent with the proposed sum rule. The original parametrization Ref. \cite{Carlson:2013xea} led to a $1.5\sigma$-disagreement because the slope parameter was obtained by an extrapolation beyond the kinematical range covered by the data without using the value of $\beta_M^d$ as a constraint. 

Another sum rule involving the $Q^2$-slope of the integrated structure functions was proposed by Bernabeu and Jarlskog \cite{Bernabeu:1973uf}. They assumed that the longitudinal amplitude obtained as a linear combination of $T_1$ and $T_2$ obeys an unsubtracted dispersion relation, and argued that the longitudinal structure function has to vanish identically at the real photon point independently of the energy to ensure gauge invariance, hence the integral becomes convergent. In this way they arrived at a sum rule for the electric polarizability $\alpha_E$ alone, which is however incompatible with the $\beta_M$ sum rule proposed here. I believe that the reason for the disagreement lies in their use of unsubtracted dispersion relation. 
Since it is the $Q^2$-slope that gives the sum rule, one in reality explicitly departs from the real photon point; then, the argument of vanishing of the longitudinal structure function at infinity is no longer valid, and one is left with a divergent integral,  so that the limit $Q^2\to0$ does not exist. 

The parametrization of deuteron quasi elastic data was used in Ref. \cite{Carlson:2013xea} to estimate the two-photon exchange correction to the $2P-2S$ Lamb shift in muonic deuterium atom. A modification of the data parametrization proposed above based on the new sum rule will lead to a different prediction for that correction. Moreover, the photonuclear sum rule discussed above can further be extended beyond its value and slope at $Q^2=0$ (TRKLB and the $\beta_M^{nucl}$ sum rule, respectively) to predict the full $Q^2$-dependence of the subtraction function via
\beqn
&&T_1^{np}(0,Q^2)-T_1^{np}(0,0)\\
&&\;\;\;=\frac{2\alpha_{em}}{M_T}\int\limits_{\nu_{min}(Q^2)}^{\nu_{max}(Q^2)}\frac{d\nu}{\nu}\left[F_1(\nu,Q^2)-F_1(\nu,0)\right],\nn
\eeqn
which contributes to the shift of the $2S$ state through
\beqn
\Delta E_{2S}^{Sub}=4\alpha_{em}\phi^2_{2S}(0)\!\!\int\limits_0^\infty\!\! dQ\gamma_1(\tau_l)\frac{T_1^{np}(0,Q^2)-T_1^{np}(0,0)}{Q^2},\nn\\
\eeqn
with $\gamma_1(x)=(1-2x)\sqrt{1+x}+2x^{3/2}$, $\tau_l=Q^2/(4m_l^2)$, $m_l$ the lepton mass, and 
$\phi_{nS}^2(0)=(Z\alpha_{em}m_r)^3/\pi n^3$ the squared atomic wave function at origin with the reduced mass $m_r=M_Tm_l/(M_T+m_l)$. The value of $T_1(0,0)$ is subtracted to account for its inclusion in the lowest order atomic calculation. A similar approach based on the finite energy sum rule obtained upon removing the Regge-behaved part of the hadronic photoabsorption, was applied to the muonic hydrogen Lamb shift  \cite{Gorchtein:2013yga}. 

\begin{table}
  \begin{tabular}{c|c|c|c}
\hline
$\Delta E_{2S}^{i}$ & This work & Ref. \cite{Carlson:2013xea}  &  Refs. \cite{Pachucki:2011xr,Hernandez:2014pwa,Pachucki:2015uga,Krauth:2015nja}\\
\hline
\hline
$\Delta E_{2S}^{inel}$ &  $-2.294(740)$  &  $-2.357(740)$  & --\\
\hline
$\Delta E_{2S}^{Subt}$ &  $0.505(35)(40)$  &  $0.763(40)$ & --\\
\hline
${\Delta E_{2S}^{Tot}} $ &  $-1.945(740)$ $^*$ &  $-1.750(740)$ $^*$ & $-1.709(15)$ \\
\hline
 \end{tabular}
\caption{TPE contributions to the shift of the $2S$ state in muonic deuterium in units of meV. The inelastic contribution is a sum of ``PWIA", ``FSI", ``$\perp$" and ``hadr" contributions listed in Table I of \cite{Carlson:2013xea}. The numbers in the first and second column in this row correspond to the use of $\tilde f_T^{FSI}$ and $f_T^{FSI}$, respectively. Subtraction contribution is calculated with the sum rule in this work, while the number in the second column is a sum of ``Th." and ``$\beta$" terms in Table I of \cite{Carlson:2013xea}. The total contribution is obtained by adding the upper two numbers with elastic term obtained in \cite{Carlson:2013xea}, and the star indicates the inclusion of the internal Coulomb correction of $0.261$ meV \cite{Krauth:2015nja}. Total contribution summarizing potential models calculations \cite{Krauth:2015nja} is listed in the rightmost column.}
\label{tab1}
\end{table}

Numerical evaluation leads to a new prediction (including the inelastic Coulomb correction, as pointed out in Ref. \cite{Krauth:2015nja}) $\Delta E_{2S}^{Tot}=-1.945(740)(40)$ meV, to be compared to $\Delta E_{2S}^{Tot}=-1.750(740)$ meV \cite{Carlson:2013xea}. The two values agree within the error that is dominated by the uncertainty due to the low-$Q^2$ behavior of the quasi elastic cross sections. The systematical uncertainty in the second bracket is due to the use of the sum rule for the subtraction term, and was estimated by varying the value of $\nu_{max}$ between 30 MeV above the quasi elastic peak, and the pion production threshold. An additional 0.01 meV uncertainty due to $\beta_M^{p,n}$  was added in quadrature. It amounts in $\approx8\%$ uncertainty and can be compared to 1\% in the sum rule for $\beta_M^d$. The reason for the larger uncertainty is mostly in a steep rise with $Q^2$ of the QE peak that resides at higher energy than the threshold peak that completely dominates at $Q^2=0$.

The large uncertainty of the DR result at present prevents one from talking of a disagreement between the new prediction and other models, nevertheless when new deuteron quasi elastic data at lower $Q^2$ will become available \cite{Distler} the uncertainty may be sizably reduced \cite{Carlson:2013xea}. In that case the shift of $-0.195$ meV will result in a different value of the deuteron charge radius extracted from the $\mu D$ Lamb shift measurement. Using $\Delta E_{2S}^{R_d}=6.1103(3)(R_d/{\rm fm})^2$ meV \cite{Krauth:2015nja}, the extracted value of $R_d$ would be larger by $\delta R_d=0.007$ fm. It is smaller than the uncertainty of the radius extraction from scattering data $R_d^{e-D}=2.128(11)$ fm but considerably larger than that using the isotope shift measurements \cite{Huber:1998zz,Parthey:2010aya} and muonic hydrogen Lamb shift \cite{Pohl:2010zza,Antognini:1900ns}, as well as the expected uncertainty of the muonic deuterium data. The method based on the new sum rule provides a different basis for estimating the subtraction function, as compared to the minimalist assumption used in Ref.  \cite{Carlson:2013xea} that the $Q^2$-dependence of the deuteron magnetic polarizability resembles that of the charge form factor $\beta_M^d(Q^2)\sim\beta_M^d F_C^d(Q^2)$. The sum rule-based calculation can be seen as a valuable systematic study of DR calculations. A direct calculation of $\beta_M^d(Q^2)$, e.g., in an EFT approach would help further assessing this systematics.

The method proposed here can be used for calculating the subtraction function contribution to the Lamb shift in other light muonic atoms with the new experiments underway \cite{muHe}. For nuclei beyond deuteron a reliable estimate of $\beta_M^{nucl}$ in potential models and in effective theories might be considerably more complicated. The proposed sum rule may serve a model-independent tool to extract $\beta_M^{nucl}$ from data, {\it e.g.} interpret measurements of M1 strength in heavy nuclei \cite{Tamii:2011pv,Matsubara:2015tua}.

Currently, models of a strongly bound composite Dark Matter (DM) \cite{Kribs:2009fy} have received much attention. Such DM particles would have electromagnetic polarizabilities and could interact with ordinary matter by means of the two-photon exchange \cite{Appelquist:2015zfa}. At present, estimates of the nuclear part of the interaction have a modest $\pm$ order of magnitude accuracy \cite{Appelquist:2015zfa}. For more quantitative calculations based on dispersion relations the new sum rule will help constraining the subtraction function contribution. 

In summary, I proposed a new sum rule that generalizes the Levinger-Bethe sum rule to the case of virtual photons. Its slope at zero photon virtuality relates the nuclear magnetic polarizability to the slope of the  transverse photoabsorption cross section integrated over the nuclear energy range. I showed that the quasielastic data on the deuteron are compatible with the sum rule, and applied its full version to the calculation of the Lamb shift in muonic deuterium. I discussed applications to light muonic atoms and direct DM detection. 

My gratitude goes to M. Birse for detailed and encouraging discussions during and after his short visit to Mainz. I furthermore acknowledge suggestions and critique by C. E. Carlson, V. Pascalutsa and M. Vanderhaeghen, and the support by the Deutsche Forschungsgemeinshaft through the Collaborative Research Center ``The Low-Energy Frontier of the Standard Model" CRC 1044.


%

%

\begin{thebibliography}{99}

\bibitem{Kronig} R.~Kronig, J.~Opt.~Soc.~Am. {\bf 12}, 547 (1926).

\bibitem{Kramers} H.~A.~Kramers, Atti.~congr.~intern.~fis. Como {\bf 2}, 545 (1927).

\bibitem{Thomas} W. Thomas, Naturwissenschaften 13, 627 (1925).

\bibitem{Reiche} F. Reiche and W. Thomas, Z. Phys. 34, 510 (1925).

\bibitem{Kuhn} W. Kuhn, Z. Phys. 33, 408 (1925).

\bibitem{Levinger} J. S. Levinger and H. A. Bethe Phys. Rev. 78, 115 (1950).

\bibitem{GellMann:1954db}
  M.~Gell-Mann, M.~L.~Goldberger and W.~E.~Thirring,
  Phys.\ Rev.\  {\bf 95} (1954) 1612.
  
\bibitem{Gorchtein:2011xx}
  M.~Gorchtein, T.~Hobbs, J.~T.~Londergan and A.~P.~Szczepaniak,
  Phys.\ Rev.\ C {\bf 84} (2011) 065202
  
  

\bibitem{Bloom:1970xb}
  E.~D.~Bloom and F.~J.~Gilman,
  Phys.\ Rev.\ Lett.\ {\bf 25} (1970) 1140.
  
\bibitem{Birse:2012eb}
  M.~C.~Birse and J.~A.~McGovern,
  Eur.\ Phys.\ J.\ A {\bf 48} (2012) 120
  
\bibitem{Harvey:1964}  R.~R.~Harvey, J.~T.~Caldwell, R.~L.~Bramblett, and S.~C.~Fultz,
Phys.\ Rev.\ {\bf 138}, B126 (1964).
  
\bibitem{Hesse:1970cy}
  W.~P.~Hesse, D.~O.~Caldwell, V.~B.~Elings, R.~J.~Morrison, F.~V.~Murphy, B.~W.~Worster and D.~E.~Yount,
  Phys.\ Rev.\ Lett.\  {\bf 25} (1970) 613.
  
\bibitem{Caldwell:1973bu}
  D.~O.~Caldwell, V.~B.~Elings, W.~P.~Hesse, R.~J.~Morrison, F.~V.~Murphy and D.~E.~Yount,
  Phys.\ Rev.\ D {\bf 7} (1973) 1362.
  
\bibitem{Caldwell:1978ik}
  D.~O.~Caldwell {\it et al.},
  Phys.\ Rev.\ Lett.\  {\bf 42} (1979) 553.

\bibitem{Bianchi:1995vb}
  N.~Bianchi {\it et al.},
  Phys.\ Rev.\ C {\bf 54} (1996) 1688.

\bibitem{BermanFultz1975}
B.~L.~Berman, S.~C.~Fultz, Rev.\ Mod.\ Phys.\ Vol.~{\bf 47}, No.~{\bf  3}, 713 (1975).

\bibitem{Carlson:2013xea}
  C.~E.~Carlson, M.~Gorchtein and M.~Vanderhaeghen,
  Phys.\ Rev.\ A {\bf 89} (2014) 2,  022504
  
  
\bibitem{Baldin} A.~M.~Baldin, Nucl.~Phys. {\bf 18}, 310 (1960).

\bibitem{Chen:1998vi}
  J.~W.~Chen, H.~W.~Griesshammer, M.~J.~Savage and R.~P.~Springer,
  Nucl.\ Phys.\ A {\bf 644} (1998) 221
  
  \bibitem{Friar} J.~L.~Friar, S.~Fallieros, E.~L.~Tomusiak, D.~Skopik, and E.~G.~Fuller, Phys.~Rev.~C {\bf 27}, 1364(R) (1983).

\bibitem{Pohl:2010zza}
  R.~Pohl {\it et al.},
  Nature {\bf 466} (2010) 213.
   
\bibitem{Antognini:1900ns}
  A.~Antognini {\it et al.},
  Science {\bf 339} (2013) 417.
  
\bibitem{PDG}
  K.~A.~Olive {\it et al.}~(Particle Data Group), Chin.~Phys.~C {\bf 38}, 090001 (2014).

\bibitem{Mohr:2012tt}
  P.~J.~Mohr, B.~N.~Taylor and D.~B.~Newell,
  Rev.\ Mod.\ Phys.\  {\bf 84} (2012) 1527
  
\bibitem{Bernabeu:1973uf}
  J.~Bernabeu, C.~Jarlskog,
  Nucl.\ Phys.\ B {\bf 75} (1974) 59.
  
\bibitem{Gorchtein:2013yga}
  M.~Gorchtein, F.~J.~Llanes-Estrada and A.~P.~Szczepaniak,
  Phys.\ Rev.\ A {\bf 87} (2013) 5,  052501
  
\bibitem{Krauth:2015nja}
  J.~J.~Krauth 
{\it et al.},  arXiv:1506.01298 
 
\bibitem{Pachucki:2011xr}
  K.~Pachucki,
  Phys.\ Rev.\ Lett.\  {\bf 106} (2011) 193007

\bibitem{Hernandez:2014pwa}
  O.~J.~Hernandez, C.~Ji, S.~Bacca, N.~N.~Dinur and N.~Barnea,
  Phys.\ Lett.\ B {\bf 736} (2014) 344
   
\bibitem{Pachucki:2015uga}
  K.~Pachucki and A.~Wienczek,
  Phys.\ Rev.\ A {\bf 91} (2015) 4,  040503
 
 \bibitem{Distler} M. Distler and J. Bernauer (private communication).
 
\bibitem{Huber:1998zz}
  A.~Huber 
 {\it et al.},  Phys.\ Rev.\ Lett.\  {\bf 80} (1998) 468.
 
\bibitem{Parthey:2010aya}
  C.~G.~Parthey 
 {\it et al.},   Phys.\ Rev.\ Lett.\  {\bf 104} (2010) 233001.
 
 \bibitem{muHe} A.~Antognini, F.~Kottmann and R.~Pohl, private communication.
 
\bibitem{Tamii:2011pv}
  A.~Tamii {\it et al.},
  Phys.\ Rev.\ Lett.\  {\bf 107} (2011) 062502
  [arXiv:1104.5431 [nucl-ex]].
  
\bibitem{Matsubara:2015tua}
  H.~Matsubara {\it et al.},
  Phys.\ Rev.\ Lett.\  {\bf 115} (2015) 10,  102501.
 
\bibitem{Kribs:2009fy}
  G.~D.~Kribs, T.~S.~Roy, J.~Terning and K.~M.~Zurek,
  Phys.\ Rev.\ D {\bf 81} (2010) 095001.
  
\bibitem{Appelquist:2015zfa}
  T.~Appelquist {\it et al.},
  arXiv:1503.04205 [hep-ph].
  
\end{thebibliography}
\end{document}